\documentclass[iop]{emulateapj}

\usepackage{natbib}
\newcommand{\twco}{$^{12}$CO(1--0)}
\newcommand{\thco}{$^{13}$CO(1--0)}
\newcommand{\Ha}{H$\alpha$}
\newcommand{\Pa}{Pa$\alpha$}
\newcommand{\Msun}{$M_\odot$}

\newcommand{\kms}{km s$^{-1}$}

\newcommand{\xco}{$X_{\rm CO}$}
\newcommand{\xcouni}{cm$^{-2}$ (K km s$^{-1}$)$^{-1}$}

\shorttitle{Molecular Gas Evolution in M 51}
\shortauthors{Egusa, Koda, \& Scoville}

\begin{document}


\title{Molecular Gas Evolution across a Spiral Arm in M 51}


\author{Fumi Egusa\altaffilmark{1,3}, 
Jin Koda\altaffilmark{2}, and Nick Scoville\altaffilmark{1}}

\email{fegusa@ir.isas.jaxa.jp}




\altaffiltext{1}{California Institute of Technology, MS 249-17 Pasadena, CA 91125, USA}
\altaffiltext{2}{SUNY Stony Brook, Stony Brook, NY 11794, USA}
\altaffiltext{3}{Current address: Institute of Space and Astronautical Science, 
Japan Aerospace Exploration Agency, Sagamihara, Kanagawa 252-5210, Japan}


\begin{abstract}
 We present sensitive and high angular resolution CO(1--0) data obtained by 
the Combined Array for Research in Millimeter-wave Astronomy (CARMA) observations 
toward the nearby grand-design spiral galaxy M 51. 
 The angular resolution of $0.7''$ corresponds to 30 pc, which is similar to the typical size 
of Giant Molecular Clouds (GMCs), and the sensitivity is also high enough to detect typical GMCs.
 Within the $1'$ field of view centered on a spiral arm, a number of GMC-scale structures 
 are detected as clumps.
 However, only a few clumps are found to be associated with each 
Giant Molecular Association (GMA), and 
more than 90\% of the total flux is resolved out in our data. 
 Considering the high sensitivity and resolution of our data, 
these results indicate that GMAs are not mere confusion of GMCs 
but plausibly smooth structures.
 In addition, we have found that the most massive clumps are located 
downstream of the spiral arm, 
which suggests that they are at a later stage of molecular cloud evolution across the arm 
and plausibly are cores of GMAs.
 By comparing with \Ha ~and \Pa ~images, most of these cores 
are found to have nearby star forming regions.
 We thus propose an evolutionary scenario for the interstellar medium, 
in which smaller molecular clouds collide to form smooth GMAs at 
spiral arm regions and then star formation is triggered in 
the GMA cores.
 Our new CO data have revealed the internal structure of GMAs at GMC scales, 
finding the most massive substructures on the downstream side of the arm in 
close association with the brightest \ion{H}{2} regions. 
\end{abstract}




\section{Introduction} \label{sec:intro}
 As all the stars are formed from the interstellar medium (ISM), 
the evolution of physical and chemical conditions of ISM is an important 
clue to understand star formation, its history, and galaxy formation.
 Spiral galaxies have been studied often with this goal; 
they are actively forming stars in spiral arms
and the sequence of the ISM evolution from molecular gas to young stars 
can be investigated across the arms.
 There have been many discussions on the nature of the spiral structure.
 \citet{LS64} described it as a quasi-stationary structure sustained by density waves, 
while tidal waves due to interactions with a companion galaxy or shear motions in 
the galactic disk can also create transient spiral structures \citep[e.g.,][]{TT72}.

 With its proximity, nearly face-on disk, and grand-design spiral structure, 
the Whirlpool galaxy, \object[NGC 5194]{M 51} or NGC 5194, 
has been a popular target for both observational and theoretical studies.
 Strong streaming motions found in the two spiral arms \citep{Aal99,Shet07} 
indicate the existence of density waves in this galaxy, while it has been 
suggested that the spiral structure is not fully stable across time or radius 
\citep[e.g.,][]{Mei08,Dob10}.
 An interaction with the companion galaxy, NGC 5195, has been thought 
to enhance or even trigger the prominent spiral structure in the disk of M 51.
 Molecular gas in this grand-design spiral galaxy 
traced by dust lanes and CO emission lines has been found upstream
of star forming regions traced by \Ha ~\citep[e.g.,][]{Vog88,RK90}.
 \citet{Til88} found similar spatial displacement between non-thermal and thermal 
radio continuum; spiral arms seen in the non-thermal component are coincident 
with dust lanes while most of the thermal component peaks agree with \ion{H}{2} 
regions.
 These offsets, also found in other spiral galaxies, 
are consistent with the density wave theory and represents 
the timescale for molecular clouds to form stars \citep[e.g.,][]{Rob69}. 
 Variations of physical properties across the arms thus correspond to 
the evolution of the ISM in spiral galaxies.
 \citet{Egu09} utilized offsets between CO and \Ha ~to derive a star formation timescale 
and pattern speed in nearby spiral galaxies.

 Given that the typical width of spiral arms is about 500 pc, investigating 
the ISM evolution across the arms requires images with a high 
spatial resolution, which has been difficult especially for radio observations.
 Instead, comparison between arm and interarm regions 
has been used to diagnose how the spiral structure affects the ISM properties.
%
 From the \twco ~data covering the entire disk of M 51 taken with 
the Combined Array for Research in Millimeter-wave Astronomy (CARMA) 
and Nobeyama 45m telescope, \citet[hereafter K09]{Koda09} found that 
Giant Molecular Associations (GMAs; $\gtrsim 10^7~M_\odot$; \citealt{Vog88}) are only in arm regions 
while Giant Molecular Clouds (GMCs; $10^{5-6}~M_\odot$; \citealt{SSS79IAU}) 
are in both arm and interarm regions.
 On the other hand, bright young stars delineate narrow spiral arms 
while older stars are distributed more smoothly over a galactic disk \citep[e.g.,][]{San61}. 
 For M 51, \citet{Sco01} cataloged \ion{H}{2} regions from a high-resolution \Ha ~image 
taken with the Hubble Space Telescope (HST) and found that about half of them are 
confined to the arm region which occupies only quarter of the disk area.
 It is thus expected that GMAs play a critical role in 
massive star formation in the spiral arms.

 However, the typical size of GMAs and \ion{H}{2} regions are quite different; 
$\gtrsim 200$ pc and $\sim 30$ pc, respectively.
 Thus, the key to understanding the initiation of massive OB star clusters
is the internal structure of GMAs.
 K09 measured the surface density of GMAs to be similar to that of typical GMCs
and suggested that GMAs are not simply overlapping of GMCs but are 
distinct structures where the entire volume is virtually filled with molecular gas.
 Nevertheless, the spatial resolution of their data ($\sim 200$ pc) 
is not high enough to resolve GMAs. 
 Since previous CO observations with high spatial resolution similar to 
the size of GMCs and \ion{H}{2} regions have only been done for the central region of galaxies 
or for dwarf and 
flocculent spiral galaxies in the Local Group, the internal structure of GMAs 
in grand-design spiral galaxies has not been investigated before.

 To resolve GMAs and detect GMC-scale structures ($\sim 40$ pc) in spiral arms, 
we have carried out new CARMA observations in the CO lines 
toward M 51 with a very high angular resolution ($\sim 0.7''$).
 In this paper, we present results from our new CO observations 
and discuss the internal structure of GMAs and the relationship of the brightest substructures to 
the spiral structure and star forming regions.
 At the adopted distance of M 51 \citep[8.4 Mpc;][]{Fel97}, 
$1''$ corresponds to 41 pc. 

\section{Observations and Data Reduction} \label{sec:obs}

 The CO(1--0) line observations were carried out 
at CARMA in November through December 2008. 
 This array has six 10 m dishes and 
nine 6 m dishes and thus provides 105 baselines.
 The full width at half maximum (FWHM) of the primary beam 
is about $60''$ and $100''$ 
for the 10 m and 6 m dishes, respectively. 
 The B-array configuration, whose typical angular resolution is $0.8''$ at 3 mm, 
was selected to resolve GMAs and detect GMC-scale structures in M 51.
 A single pointing was centered on 
the brightest GMA in the K09 data located 
about $30''$ from the galactic center in the south-west direction.
 The pointing center coordinate is (RA, DEC)=(13:29:50.16, +47:11:28.39) in J2000.
 We used the hybrid correlator mode (three 62 MHz bands for the source and 
three 500 MHz bands for the calibrator) in order to achieve high spectral resolution 
for the source and high sensitivity for the calibrator.
 Channel width for the source is 0.977 MHz or 2.54 \kms ~at the frequency of 115 GHz.
 The intermediate frequency (IF) was set to include the \twco ~line in the upper side band (USB) 
and \thco ~in the lower side band (LSB).
 Data were taken through the standard cal-source cycle with 5 minutes integration on 
the phase calibrator, 1153+495, which is 16 degrees away from the source, 
and 15 minutes on source.
 The passband observations were done at the beginning of each track with a brighter source, 
and the data were used to calibrate phase and amplitude offsets between narrow and wide bands 
as well as the normal passband calibration.
 The total observing time is about 34 hours with five tracks.
 In terms of the weather condition, the phase rms was $\sim 150$--$200 ~\mu$m and 
the opacity at 230 GHz was $\sim 0.3$, which is typical for 3 mm observations 
with the B-array configuration.

 Data reduction was done with the Multichannel Image Reconstruction, Image Analysis and Display 
software \citep[MIRIAD;][]{Sau95}.
 The flux of the phase calibrator was derived to be 0.81 Jy at 114.75 GHz (USB) 
and 0.74 Jy at 110.02 GHz (LSB) 
from the comparison with bright quasars 
whose flux had been derived from other observations with planets.
 The uncertainty in the flux calibration is estimated to be 15\% ($\sim 0.1$ Jy).
 The synthesized beam from the natural weighting is $0.73''\times 0.60''$ for USB 
and $0.77''\times 0.62''$ for LSB, 
corresponding to $\sim 31$ pc$\times 25$ pc at the distance of M 51. 
 The noise rms ($\sigma$) measured at central $40''\times 40''$ in emission-free 
channels of the dirty map 
is 12 mJy beam$^{-1}$ or 2.5 K for USB 
and 7.2 mJy beam$^{-1}$ or 1.5 K for LSB, 
with the channel width of 5.08 \kms ~as Hanning smoothing was applied. 
 The baseline length ranges 23.9 k$\lambda$ to 362 k$\lambda$, 
which corresponds to $8.6''$ to $0.60''$ in angular resolution or 
360 pc to 24 pc.
 Given that the typical width of molecular spiral arm seen in the K09 CO data 
is $\sim 10''$ or 400 pc, most of the structures smaller than the arm 
down to 24 pc should be detected.

 The dirty map was CLEANed by the use of the MIRIAD task {\tt mossdi} 
with the flux cutoff of $1\sigma$.
 In order to shorten the time needed to CLEAN, we applied a mask 
made by the combination of (i) the field of view, 
and (ii) mask from the K09 cube data.
 First, we created the sensitivity map from the primary beam patterns by the task {\tt mossen} and 
defined the field of view as where the sensitivity is lower than 
1.5 times the one at the pointing center.
 The diameter of this field of view is $53.25''$ and pixels outside have been masked out 
to avoid confusion with elevated noise.
 We made the RA-DEC, RA-VEL, and DEC-VEL projections of the K09 cube 
and manually drew a polygon around emissions to make the mask 
for the entire galaxy.
  Within the field of view, only the region inside the K09 mask 
has been used for CLEANing.
 We refer to this CLEANed dataset as B-array data hereafter.

\begin{figure*}[t!]
\begin{minipage}[c]{0.6\linewidth}
\includegraphics[width=\linewidth]{f1a.ps}
\end{minipage}
\hspace{12pt}
\begin{minipage}[c]{0.35\linewidth}
\includegraphics[width=\linewidth]{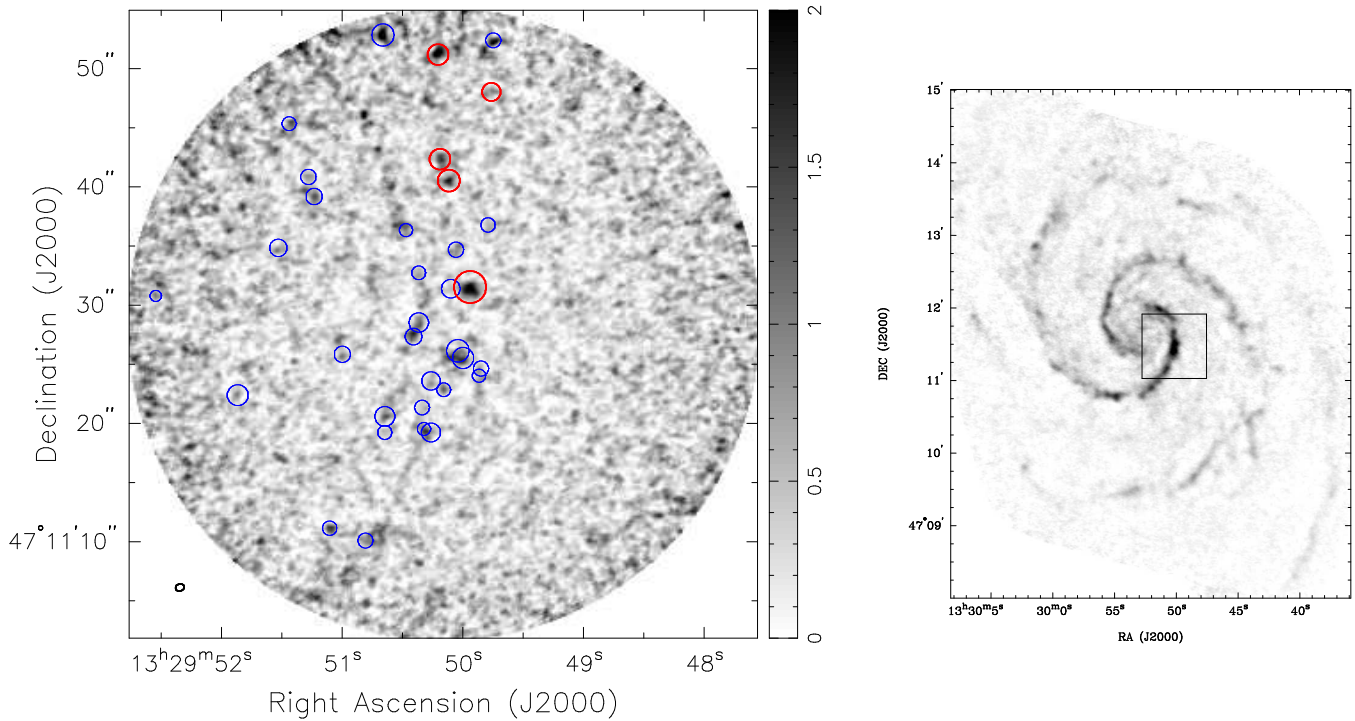}
\end{minipage}
\caption{{\bf Left:} Integrated intensity map of \twco ~data from the B-array CARMA observations. 
The unit of the gray scale is Jy beam$^{-1}$ \kms.
The synthesized beam ($0.73''\times 0.60''$) is shown as the ellipse on the bottom left corner.
Circles superimposed on the gray-scale image 
indicate clumps detected by the clumpfind, with the line width and color corresponding 
to the local S/N of the peak flux; thin blue is for S/N $\geq 4$ and thick red is for S/N $\geq 5$. 
See \S \ref{sec:clfind} for detail.
{\bf Right:} The field of view of the left image is shown as the box overlaid on 
the lower resolution \twco ~integrated intensity map from K09. 
Note that the color scale for the two images is not the same.}
\label{fig:mom0}
\end{figure*}

 The integrated intensity or moment 0 ($I_{\rm CO}=\int T_{\rm CO}dv$) map 
was created by integrating the B-array data cube 
with the velocity range of 385.5--563.2 \kms. 
 The flux cutoff for making the $I_{\rm CO}$ map was set to be 
proportional to the local sensitivity; $1\sigma$ at the pointing center 
and $1.5\sigma$ at the edge of the field of view.
 The resulting $I_{\rm CO}$ map for the \twco ~data is shown in 
the left panel of Figure \ref{fig:mom0}, while the B-array field of view is indicated 
as a box on the K09 $I_{\rm CO}$ map in the right panel.
 
\section{Results and discussion} \label{sec:res}

\subsection{Detecting GMC-scale structures}\label{sec:clfind}
 In order to identify each molecular cloud structure, 
the clumpfind method provided by \citet{clfind} has been applied to 
the CLEANed \twco ~data cube with the lowest level and contouring step of $1\sigma$.
 Both of these two parameters are constant within the field of view.
 We have then defined detections as clumps (i) within the field of view, 
(ii) with the peak flux whose {\it local} signal-to-noise ratio ${\rm (S/N)} \geq 4$, 
and (iii) with the FWHM in velocity larger than 0.5 pixel (or 2.54 \kms).
 We have calculated this local S/N, 
considering the variable sensitivity within the field of view.
 
 The selected 34 clumps are indicated by circles in 
the left panel of Figure \ref{fig:mom0} superimposed on the moment 0 map.
 The line width and color represents the local S/N of each clump; 
thin blue is for $5 >$ S/N $\geq 4$ while thick red is for S/N $\geq 5$.
 The radius measured by the clumpfind is $0.5''$--$1.4''$ or 10--60 pc, 
corresponding to the radius of the circles in the figure.
 The FWHM in RA and DEC also fall in the same range with the radius and 
about one-third of the selected clumps are resolved.
 While most of the clumps are found in the arm,
larger and more massive clumps are located 
only on the west side of the spiral arm.
 Assuming a trailing arm, this corresponds to the downstream side.
 On the other hand, some of smaller clumps are found in the upstream interarm region 
(east side of the field of view).
 In \S \ref{sec:K09}, we discuss it in more detail 
by comparison with the K09 data.

 From Figure \ref{fig:mom0}, we have noticed that there are several 
emission peaks not selected as clumps.
 Most of them are close to the edge of the field of view and thus 
are presumably elevated noise, while some appear sequentially on 
upstream side of the spiral arm. 
 We have examined all the clumps with S/N $\geq 3$ and found that 
they are superposition of several weak- and narrow-line clumps at different velocities.
 As some of them seem to be associated with \ion{H}{2} regions (see Figure \ref{fig:USB_Pa}), 
they perhaps are real molecular clouds, but we need a higher sensitivity 
to confirm that and do not include them for further discussion.

 For the \thco ~line in the LSB data, the same procedures as the \twco ~data have been 
applied and three clumps with S/N $\geq 4$ have been detected.
 None of them, however, are found to be associated with the \twco ~clumps.
 Lowering the S/N cutoff to 3, we have detected 162 clumps and found 
two clumps with S/N $\simeq 3.6$ located near \twco ~clumps; 
one is located close to the pointing center 
and coincides with the largest and most massive \twco ~clump 
(marked as `d' in Figure \ref{fig:2mom0}), while 
the other is located close to the two \twco ~clumps at $\sim 10''$ north 
of the pointing center (`b' and `c' in Figure \ref{fig:2mom0}).
 A ratio of the peak temperatures of corresponding clumps is 
$T_{\rm peak}(^{13}{\rm CO})/T_{\rm peak}(^{12}{\rm CO})\sim 0.4$, 
while a ratio of the total flux is 
$F_{\rm tot}(^{13}{\rm CO})/F_{\rm tot}(^{12}{\rm CO})\sim 0.2$.
 The average integrated intensity ratio for the Galaxy is $1/5.5$ \citep{SSS79ApJ}, 
implying that these massive clumps detected in M 51 are in similar condition to GMCs in the Galaxy.
 As these \thco ~structures do not meet the S/N criterion of for the clumps, 
we discuss only the \twco ~results hereafter.

 The conversion factor from the \twco ~integrated intensity to the H$_2$ column density 
($N ({\rm H_2}) = X_{\rm CO}I_{\rm CO}$)  
has been measured in a number of ways.
 For the Galaxy, $X_{\rm CO} \simeq 3\times 10^{20}$ \xcouni ~was derived 
by comparing the virial mass and the \twco ~flux of GMCs \citep[e.g.,][]{Solo87, YS91}.
 More recently, smaller values ($\sim 1.8\times 10^{20}$ \xcouni) have been reported 
by comparison with the $\gamma$-ray or far infrared data \citep{SM96, Dame01, Gren05}.
 \citet{Hey09} derived GMC masses from the \thco ~emission lines 
under the local thermodynamic equilibrium 
(LTE) assumption and found that the LTE masses are factor of $\sim 5$ smaller than 
the virial masses.
 The inconsistencies in \xco ~values are due to the combination of 
uncertainties and/or invalidities of the models and assumptions introduced
and imply difficulties in the \xco ~determination.

 As for the Galaxy, 
derived \xco ~values for M 51 depend largely on data and methods.
 While the virial masses of GMAs derived by \citet{RK90} agree with 
$X_{\rm CO} = 3\times 10^{20}$ \xcouni, those derived by \citet{Adl92} based on a 
different CO dataset suggest $X_{\rm CO} = 1.2\times 10^{20}$ \xcouni.
 Even smaller values have been suggested \citep{GB93I,Rand93,NK95}.
 More recently, \citet{Schi10} applied the large velocity gradient (LVG) modeling to $2''$ resolution 
multi-transition CO data and derived $X_{\rm CO} = (1.3-1.9)\times 10^{20}$ \xcouni. 
 In this paper, we adopt $X_{\rm CO} = 1.5\times 10^{20}$ \xcouni. 
 The variations of \xco ~within the field of view 
as a function of galactocentric radius should be small enough 
to be neglected, as the data only cover $53''$ or 2.2 kpc. 

 The noise rms ($1\sigma$) in the \twco ~data thus corresponds to 
$M_{\rm H_2}=2.5\times 10^4$ \Msun ~within
the synthesized beam ($\sim 30$ pc) and one velocity channel (5.08 \kms). 
 Though the absolute mass of molecular gas could be changed by a factor of two or more 
according to adopted \xco ~values, the sensitivity in mass is still 
smaller than typical GMCs in any case.
 Furthermore, the conclusion of this paper 
does not depend on the \xco ~value, 
since we only focus on the spatial and kinematical information and discuss 
relative values of mass from the B-array and K09 data.

\subsection{Relationship to larger structures}\label{sec:K09}
\begin{figure*}
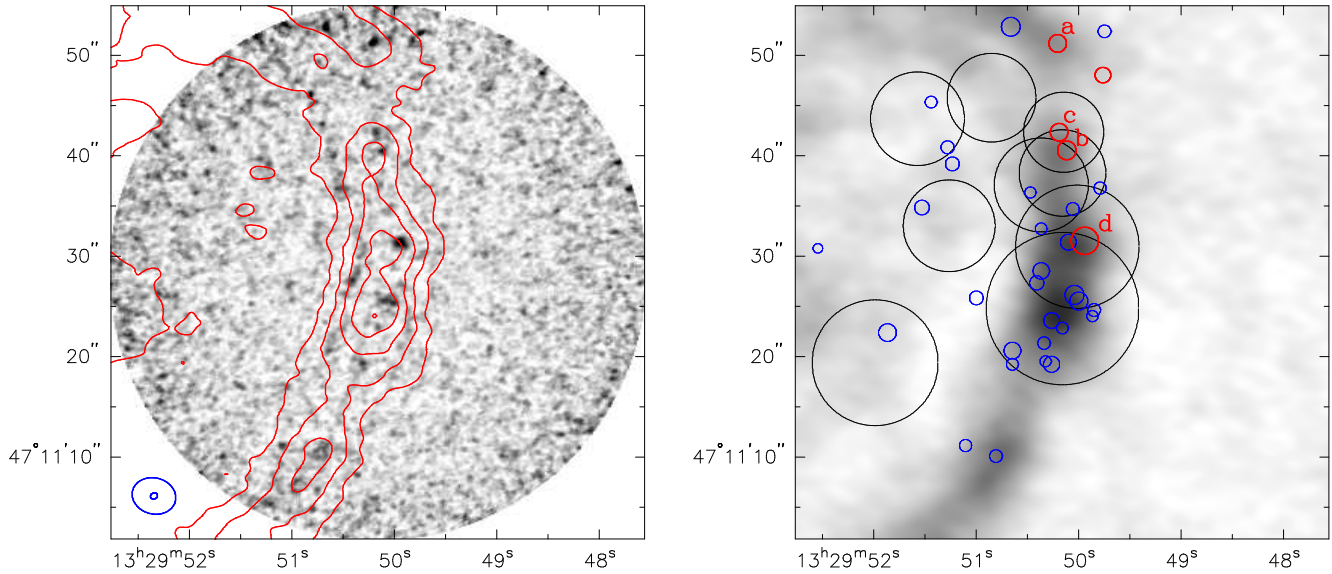

\includegraphics[width=0.47\linewidth]{f2.ps}
\hspace{12pt}
\includegraphics[width=0.47\linewidth]{f2b.ps}
\caption{{\bf Left:} Contours from the K09 moment 0 map overlaid on the B-array moment 0 map. 
The synthesized beams for these two datasets are indicated by ellipses on the bottom left corner.
{\bf Right:} Clumps from the B-array data (as circles with blue or red line) and 
GMAs from the K09 data (as circles with black lines) superimposed on the K09 moment 0 map. 
See \S \ref{sec:clfind} and \ref{sec:K09} for their definition.
Clumps whose spectra are shown in Figure \ref{fig:spec} are marked with `a'--`d'.}
\label{fig:2mom0}
\end{figure*}

 In the left panel of Figure \ref{fig:2mom0}, contours from the K09 $I_{\rm CO}$ map
are overlaid on the B-array $I_{\rm CO}$ image.
 We have applied the clumpfind algorithm to the K09 data cube and defined GMAs 
as clumps with $M > 10^7$ \Msun.
 Following the selection criteria for the B-array data, 
clumps on the data edge have been excluded.
 These selected GMAs are shown as open black circles in the right panel of Figure \ref{fig:2mom0} 
together with the detected clumps in the B-array data in the same manner as Figure \ref{fig:mom0}.
 The background image is the K09 moment 0 map.
 From this Figure, we find that GMAs are not filled with the GMC-scale clumps.
 As the $1\sigma$ sensitivity of the B-array data is about 10 times smaller 
than the typical mass of GMCs, 
we should have detected tens or more of them if GMAs were confusion of GMCs.
 This result strongly indicates that GMAs are not confusion nor overlapping of GMCs.

 The total flux within the field of view of the B-array data 
has been measured to be 67 Jy \kms ~and $1.1\times 10^3$ Jy \kms ~for the 
B-array and K09 data, respectively.
 As the K09 data include the single-dish observations with Nobeyama 45m telescope, 
it should represent the total \twco ~flux.
 The missing flux in the B-array data is thus 94\%, 
suggesting that most of the CO flux in the spiral arm of M 51 
should come from either extended structures or molecular clouds smaller than GMCs 
and supporting the aforementioned indication that GMAs are not a collection of GMCs.

\begin{figure}
\includegraphics[trim=0 0 35 0,clip,width=0.47\linewidth]{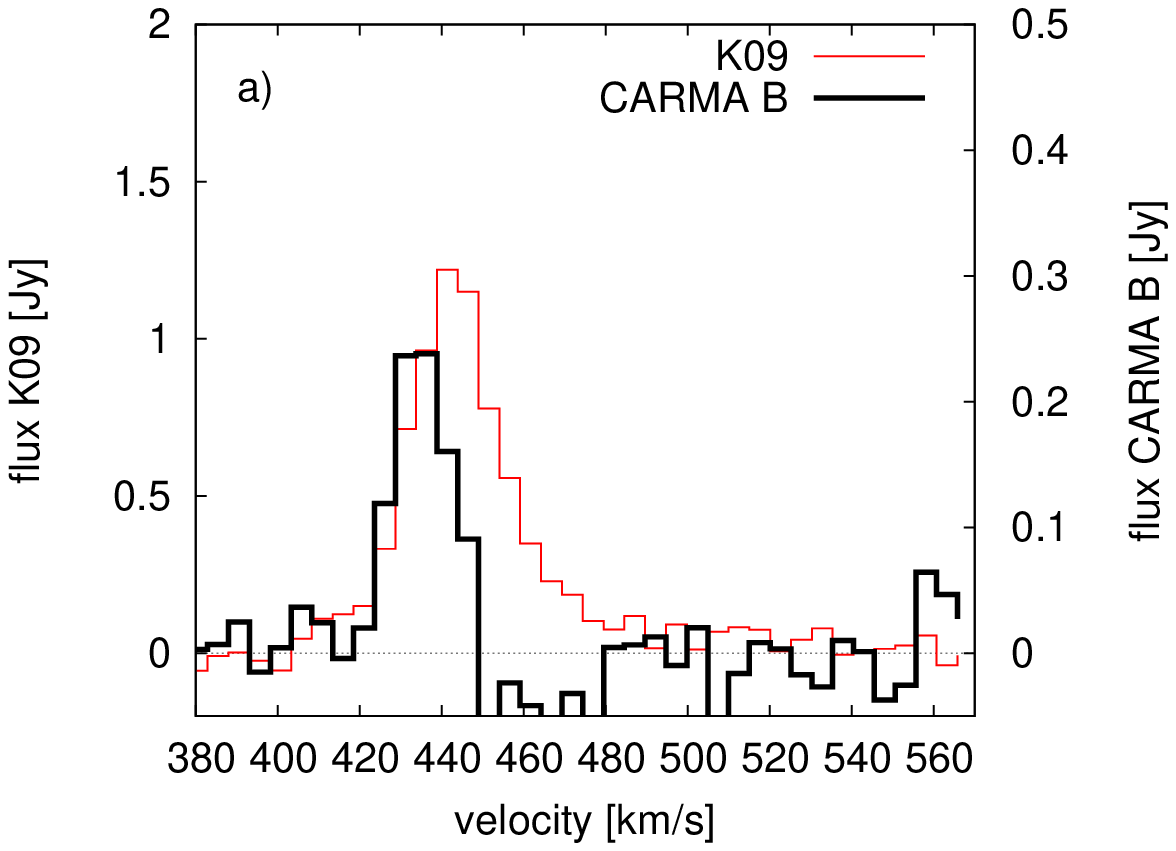}
\hspace{9pt}
\includegraphics[trim=35 0 0 0,clip,width=0.47\linewidth]{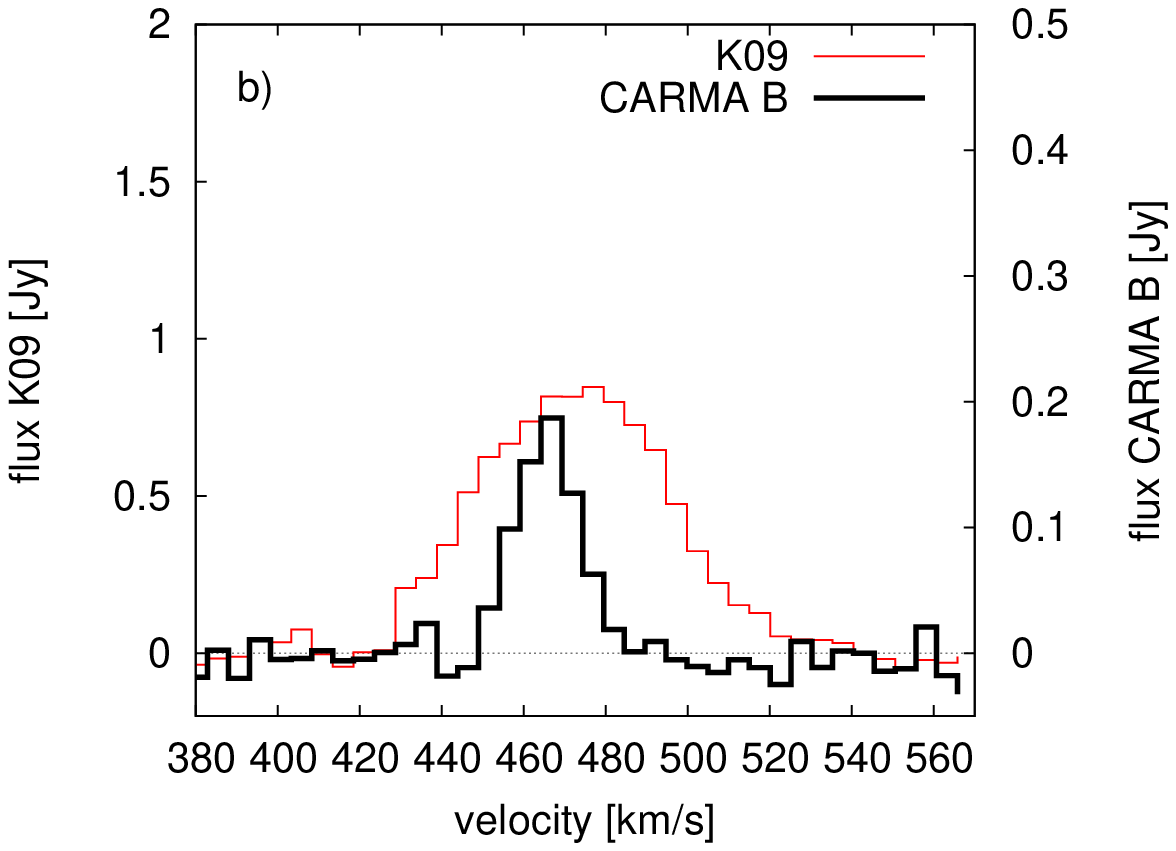}\\
\\
\includegraphics[trim=0 0 35 0,clip,width=0.47\linewidth]{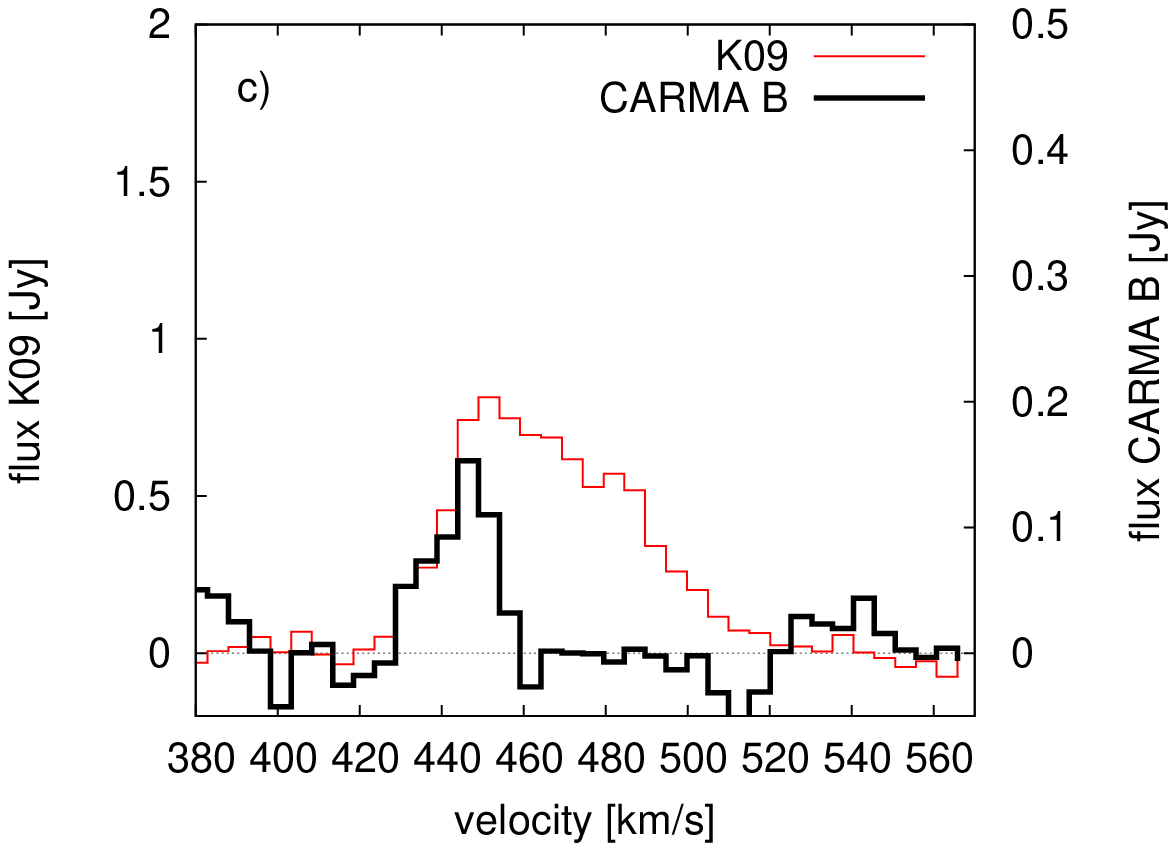}
\hspace{9pt}
\includegraphics[trim=35 0 0 0,clip,width=0.47\linewidth]{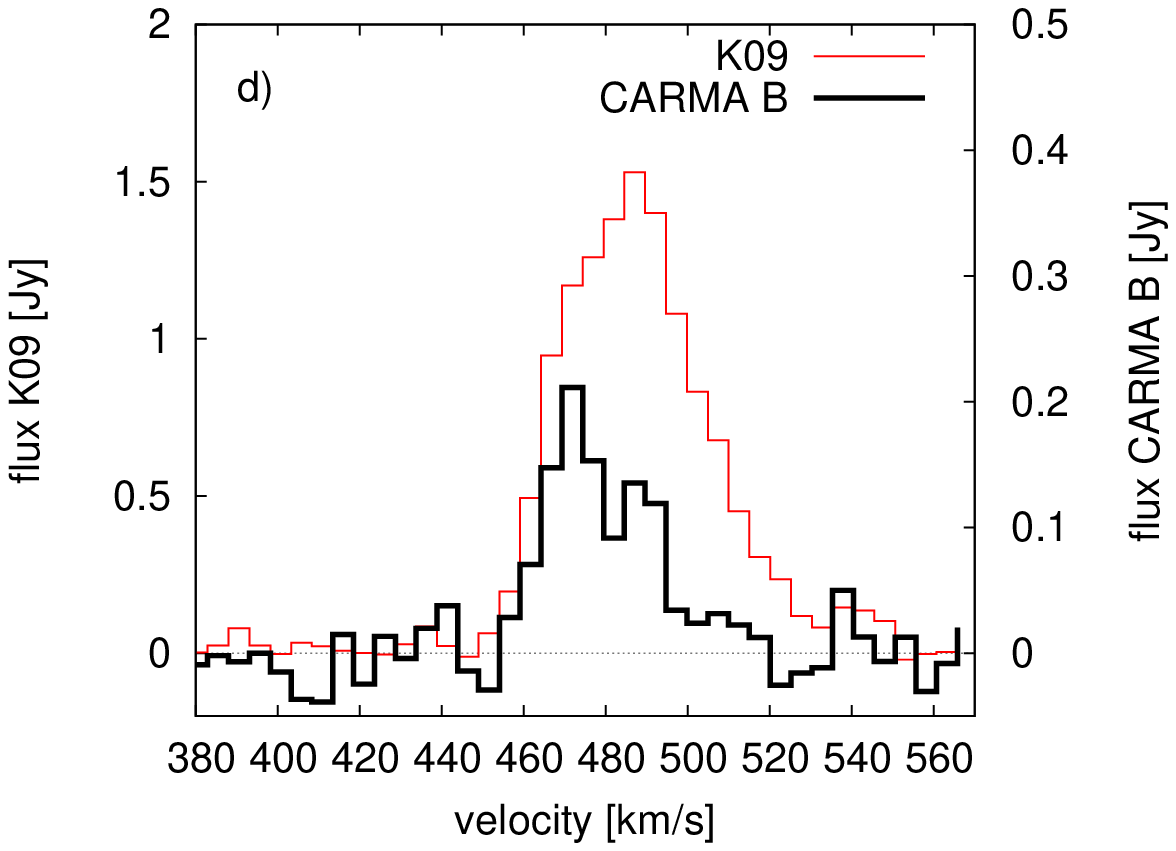}
\caption{Example of \twco ~spectra for selected massive clumps, 
which are indicated in the right panel of Figure \ref{fig:2mom0}. 
Thin red lines indicate the K09 data while black thick lines indicate the B-array data.
Note the difference in flux scale between the two datasets.
}
\label{fig:spec}
\end{figure}

\begin{figure}
\includegraphics[width=\linewidth]{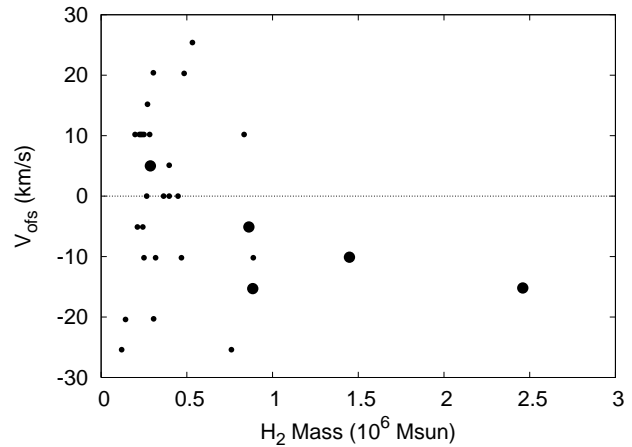}
\caption{Plot of $V_{\rm ofs}$ defined in Eq.~(\ref{eq:vofs}) v.s.\ mass of 
the selected clumps. 
Larger points represent clumps with S/N $\geq$ 5.
A few datapoints outside the $V_{\rm ofs}$ range are not shown.}
\label{fig:vofs_m}
\end{figure}

 Another important clue from comparison of the two $I_{\rm CO}$ maps 
is that their peak positions do not coincide; 
massive clumps detected in the B-array data 
are shifted a few arcseconds toward the downstream side 
of the global spiral arm seen in the K09 map.
 To investigate this difference further with the velocity information, 
we have made spectra of B-array and K09 data 
at the positions of the selected clumps.
 The box size for the average spectra was set to be the same as the diameter of 
each clump derived by the clumpfind.
 In Figure \ref{fig:spec}, 
sample spectra for massive clumps are displayed.
 The flux difference mentioned above is clearly seen also in the spectra.
 From these spectra, we have found that 
massive clumps have offsets in peak velocity between the two datasets.
 We have defined the velocity offset as 
\begin{equation}
V_{\rm ofs}=V_{\rm cl}({\rm B})-V_{\rm peak}({\rm K09}), \label{eq:vofs}
\end{equation}
where $V_{\rm cl}({\rm B})$ is the peak velocity for clumps measured by the clumpfind 
and $V_{\rm peak}({\rm K09})$ is the peak velocity in the K09 spectra.
 The $V_{\rm ofs}$ is plotted against the mass of each clump 
in Figure \ref{fig:vofs_m}.
 It is clear from this plot that almost all of the clumps with 
$M_{\rm H_2} > 5\times 10^5$ \Msun ~have negative $V_{\rm ofs}$.
 Considering the galactic rotation, negative velocities correspond to the downstream side.
 Therefore, massive clumps are kinematically as well as spatially located downstream 
of the spiral arm.
 This difference indicates that the massive clumps are in a later evolutionary stage 
compared to the larger scale structures or GMAs.
 We refer to these massive clumps as GMA cores hereafter.

 Smaller clumps are on the other hand located preferentially on upstream of the spiral arm, 
and little emission is found in the interarm region on downstream. 
 This distribution of small clumps indicates that 
they are somehow destroyed through the evolution across the spiral arm. 
 Before forming stars, the small clumps ($M \sim 10^5$ \Msun) 
could collide with each other to form larger clouds such as GMAs. 
 Meanwhile, they would be dissociated into atomic gas or broken up to 
even smaller clouds ($M < 10^5$ \Msun) after forming stars.

\subsection{Relationship to star forming regions}
 As already mentioned in \S \ref{sec:intro}, star forming regions are located 
on the downstream side of the spiral arms.
 Since we have found that GMA cores 
are also on downstream, it is worth comparing 
these two components.

\begin{figure*}[t]
\plottwo{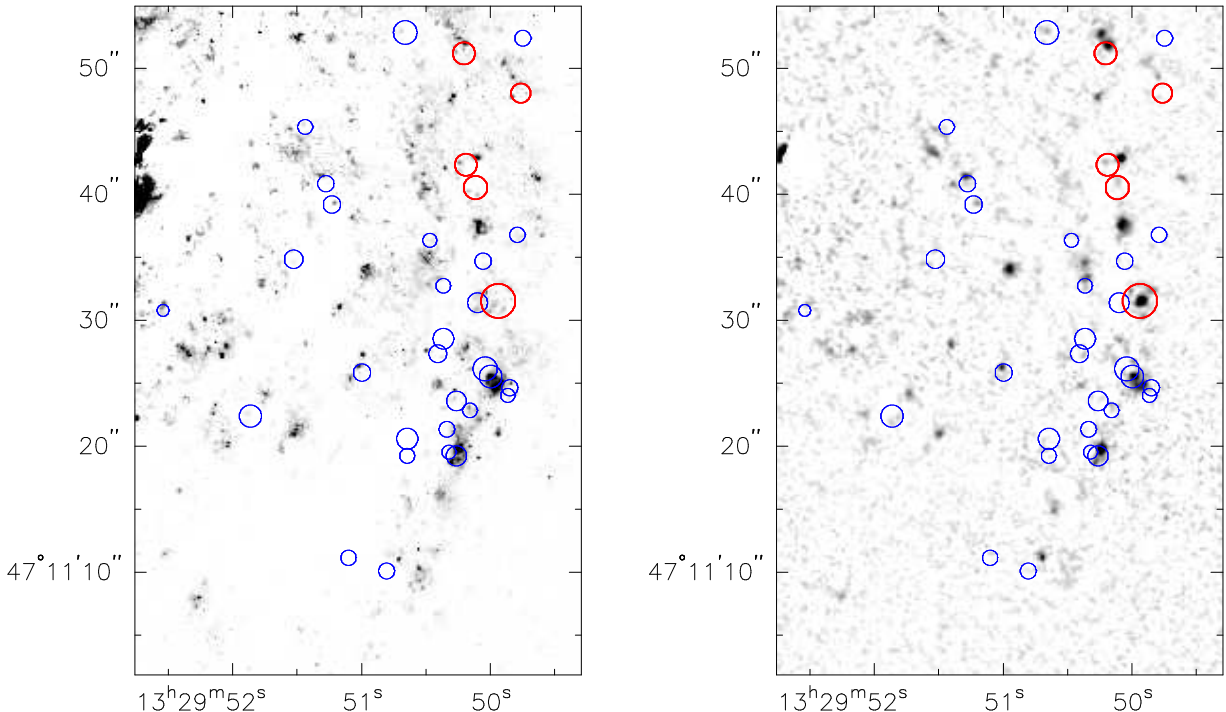}{f5b.ps}
\caption{
Selected clumps presented as circles (same as Figure \ref{fig:mom0}) 
on the HST \Ha ~(left) and \Pa ~(right) image. 
One-third from the right edge of the field of view, where no clumps have been detected, 
is not displayed.
}
\label{fig:USB_Pa}
\end{figure*}

 The \Ha ~data taken under the 
HST program (ID: 10452, PI: S.~Beckwith) 
were obtained via the HST website\footnote{http://archive.stsci.edu/prepds/m51/} 
and cover $\sim 7'\times 10'$ including 
the entire galactic disk and the companion galaxy \citep{Mut05}.
 The \Pa ~image taken for another HST program (ID: 7237, PI: N.~Scoville) 
was presented by \citet{Sco01} and covers central $3'$ of the disk.
 The astrometry of the \Pa ~image was found to be slightly 
off compared to the \Ha ~image.
 In order to match their coordinates, we measured 
the peak position of bright \ion{H}{2} regions seen in both \Ha ~and \Pa ~images, and
adjusted the coordinates of the \Pa ~image by the use of 
the task {\tt ccmap} of the Image Reduction and Analysis Facility (IRAF).\footnote{
IRAF is distributed by the National Optical Astronomy Observatories,
which are operated by the Association of Universities for Research
in Astronomy, Inc., under cooperative agreement with the National Science Foundation.}
 Although the fitting rms from {\tt ccmap} is as small as $0.17''$ and $0.13''$ 
for RA and DEC, respectively, 
the errors in the coordinates could be larger as \ion{H}{2} regions 
are extended especially in the \Ha ~image and extinction gradients might affect 
the flux distribution and thus the peak positions.
 We estimate the effective uncertainty in the coordinates to be about 40 pc or $1''$, 
which is the typical size of GMCs and \ion{H}{2} regions.
 In Figure \ref{fig:USB_Pa}, circles indicating the selected clumps 
are superimposed on  the \Ha ~and \Pa ~images.

 As expected, bright \ion{H}{2} regions are located on the downstream 
side of the CO spiral arm seen in the lower resolution K09 data.
 GMA cores found in the high resolution B-array data, on the other hand, coincide 
with them. 
 This correlation indicates that GMA cores are the site of massive star formation, 
which is consistent with the result in the previous subsection 
that they are in a later stage of the molecular gas evolution.
 Meanwhile, we have found that not all the \ion{H}{2} regions are 
accompanied by the cores, 
which implies that GMA cores 
are short-lived or are not prerequisites
for massive star formation.

 Given that only the \twco ~data are available at this resolution, 
physical conditions such as density and temperature of GMA cores 
cannot be determined uniquely; high density and high temperature 
can both produce high $I_{\rm CO}$.
 The largest core close to the pointing center, 
(marked as `d' in Figure \ref{fig:2mom0}), 
gives a clue on this issue.
 From Figure \ref{fig:USB_Pa}, we have found that it  
does not have a counterpart in \Ha ~but in \Pa, indicating a high extinction 
and thus a high column density.
 From the mass ($2.5\times 10^6$ \Msun) and radius (55 pc) 
measured by the clumpfind, the average H$_2$ column density 
is derived to be $N({\rm H_2}) = 1.6 \times 10^{22}$ cm$^{-2}$ for this core. 
 Assuming $A_V/N_{\rm H}=5.3\times 10^{-22}$ mag cm$^2$ \citep{Bohl78}, 
the estimated extinction is as high as $A_V=17$ mag. 
 The most plausible detection of \thco ~at this location 
is also consistent with the high density.
 Therefore, at least for this core, we conclude that the high $I_{\rm CO}$ 
represents high column density of molecular gas. 
 This agreement in high extinction and density also confirms 
that the coordinate adjustment for the \Pa ~image 
is correct within the core size or $\sim 1''$. 

 Additionally, it is clear from Figure \ref{fig:USB_Pa} that 
the distribution of \ion{H}{2} regions is quite similar 
in the \Ha ~and \Pa ~images, indicating only a few regions 
(such as the most massive core mentioned above) are significantly affected 
by extinction.
 This means that \Ha ~is a good tracer of the locations of current star formation 
in most cases 
and is preferable over extinction-free $24\mu$m for measuring offsets 
between molecular arm and young stellar arm \citep{Egu09} 
as it generally provides higher angular resolution.

\subsection{Evolution scenario across a spiral arm}
 The results presented above reveal the structure and distribution of 
molecular clouds at 30 pc scale for the first time in grand-design spiral galaxies.
 Based on their properties and relationship to the global (kpc) spiral structure, 
we here propose a scenario for the ISM evolution across the spiral arm.
 We should note here that smaller spatial scale across the arm corresponds 
to shorter timescale, which enables us to describe the scenario 
in unprecedented detail.

 On the upstream side of the spiral arm, 
molecular clouds of $M_{\rm H_2} \sim 10^5$ \Msun ~collide with each other 
and probably with smaller clouds to grow into GMAs ($M_{\rm H_2} \gtrsim 10^7$ \Msun).
 GMAs are presumably distinct and smooth structures and 
a few GMC-scale cores ($\sim 10^6$ \Msun) are formed within each GMA, 
which turn to be the site of OB star formation as they evolve.
 If not all the gas in cores would turn into stars, 
remaining gas in the core should break up to smaller molecular clouds 
($< 10^5$ \Msun) or become dissociated into atomic hydrogen by strong radiation from 
young stars.
 From the star formation law presented by \citet{Ken07} for M 51, 
the gas consumption timescale measured as the ratio of the total gas mass 
to the star formation rate is $\sim 10^9$ yr.
 Since the star formation timescale is much shorter \citep[$\sim 10^7$ yr;][]{Egu09} 
than this timescale, 
most of the gas in GMAs is not used by the star formation activity. 
 Such gas remains molecular through the evolutionary sequence, 
considering that the molecular fraction to the total hydrogen has been estimated as 70--80\% 
even in the interarm regions (K09).



\section{Summary}\label{sec:sum}
  We have carried out high resolution CO observations toward a spiral arm 
in the nearby spiral galaxy M 51 using the radio interferometer CARMA, in order to 
(i) resolve GMAs, (ii) detect GMC-scale structures, and (iii) understand 
their relationship to the global spiral structure. 
 For the \twco ~line, the angular resolution and noise rms (1$\sigma$) 
are $0.7''\times 0.6''$ (or $\sim 30$ pc) 
and 12 mJy beam$^{-1}$ (or 2.5 K) with the velocity width of 5.1 \kms, respectively. 
 Assuming the CO-to-H$_2$ conversion factor, 4$\sigma$ corresponds to $10^5$ \Msun ~within the beam, 
which is smaller than the mass of typical GMCs.
 The spatial resolution is similar to the typical size of GMCs, and thus 
the data quality is high enough to detect GMC-like structures.

 Within the $1'$ field of view, a number of GMC-scale structures referred as 
clumps are detected in the \twco ~data. 
 For the \thco ~line, only two marginal detections have been identified.
 By comparing with lower resolution \twco ~data (K09), we have located only a few clumps 
in each GMA and derived missing flux to be larger than 90\%.
 These results indicate that GMAs are not mere confusion of GMCs 
but are plausibly single and smooth structures.
 Among the detected clumps, we have found that the most massive 
($\gtrsim 5\times 10^5$ \Msun) clumps 
are located at the downstream side of the spiral arm spatially as well as kinematically. 
 This displacement indicates that 
they are at a later stage of the ISM evolution across the spiral arm
and plausibly are cores of GMAs.
 Spatial coincidence between the massive clumps and \ion{H}{2} regions 
supports this hypothesis.
 Smaller clumps on the other hand are found preferentially in the 
upstream side of the spiral arm.

 We thus propose an evolutionary scenario for the ISM in spiral galaxies; 
as smaller clouds in interarm regions approach to spiral arms, they cluster 
or collide to form GMAs, which are smooth and discrete structures.
 Within a GMA, a few massive and GMC-scale substructures 
referred to as cores are formed at a later stage of its lifetime.
 Star formation is triggered within the GMA cores, and finally 
the remaining gas in the core is dissociated into atomic gas or broken up to 
even smaller clouds after stars are formed.

 The sensitive and high resolution CO data from our new CARMA observations have revealed 
the internal structure of GMAs at GMC scale and its relationship to the spiral arm structure 
for the first time in external galaxies.
 This result is an important step to fully understand the ISM evolution scenario 
in spiral galaxies.



\acknowledgments

 We are grateful to E.~Schinnerer who kindly provided us 
a draft of her paper prior to publication.
 Support for CARMA construction was derived from the states of California, Illinois, and Maryland, 
the James S. McDonnell Foundation, the Gordon and Betty Moore Foundation, 
the Kenneth T. and Eileen L. Norris Foundation, the University of Chicago, 
the Associates of the California Institute of Technology, and the National Science Foundation. 
 Ongoing CARMA development and operations are supported by the National Science Foundation 
under a cooperative agreement, and by the CARMA partner universities. 
This research is partially supported by HST-AR-11261.01.

{\it Facilities:} \facility{CARMA}




\bibliography{ref}

\end{document}